\documentclass[%
 aip,
 float
 amsmath,amssymb,
 reprint,%
]{revtex4-1}

\usepackage{graphicx}
\usepackage{dcolumn}
\usepackage{bm}

\usepackage{hyperref}
\usepackage{upgreek}

\usepackage[utf8]{inputenc}
\usepackage[T1]{fontenc}
\usepackage{mathptmx}
\usepackage{etoolbox}
\usepackage{braket}	
\usepackage{amsmath}
\usepackage{comment}

\usepackage{here}
\usepackage{adjustbox}
\usepackage{siunitx}

\makeatletter
\def\@email#1#2{%
 \endgroup
 \patchcmd{\titleblock@produce}
  {\frontmatter@RRAPformat}
  {\frontmatter@RRAPformat{\produce@RRAP{*#1\href{mailto:#2}{#2}}}\frontmatter@RRAPformat}
  {}{}
}%
\makeatother

\begin{document}

\preprint{AIP/123-QED}

\title[]{Cavity-mediated collective emission from few emitters in a diamond membrane}

\author{Maximilian Pallmann$^\dag$}
\affiliation{Physikalisches Institut, Karlsruhe Institute of Technology (KIT), Wolfgang-Gaede Str. 1, 76131 Karlsruhe, Germany}

\author{Kerim Köster$^\dag$}
\affiliation{Physikalisches Institut, Karlsruhe Institute of Technology (KIT), Wolfgang-Gaede Str. 1, 76131 Karlsruhe, Germany}

\author{Yuan Zhang}
\affiliation{Institute of Quantum Materials and Physics, Henan Academy of Sciences, Zhengzhou 450046, China}
\affiliation{School of Physics and Microelectronics, Zhengzhou University, Zhengzhou 450052, China}

\author{Julia Heupel}
\affiliation{Institute of Nanostructure Technologies and Analytics (INA), Center for Interdisciplinary Nanostructure Science and Technology (CINSaT), University of Kassel, Heinrich-Plett-Straße 40, 34132 Kassel, Germany}

\author{Timon Eichhorn}
\affiliation{Physikalisches Institut, Karlsruhe Institute of Technology (KIT), Wolfgang-Gaede Str. 1, 76131 Karlsruhe, Germany}

\author{Cyril Popov}
\affiliation{Institute of Nanostructure Technologies and Analytics (INA), Center for Interdisciplinary Nanostructure Science and Technology (CINSaT), University of Kassel, Heinrich-Plett-Straße 40, 34132 Kassel, Germany}

\author{Klaus M{{\o}}lmer}
\affiliation{Niels Bohr Institute, University of Copenhagen, 2100 Copenhagen, Denmark}

\author{David Hunger}
\affiliation{Physikalisches Institut, Karlsruhe Institute of Technology (KIT), Wolfgang-Gaede Str. 1, 76131 Karlsruhe, Germany}


\begin{abstract}
$\dag$ These authors contributed equally. \newline
\newline
\noindent
When an ensemble of quantum emitters couples to a common radiation field, their polarizations can synchronize and a collective emission termed superfluorescence can occur. Entering this regime in a free-space setting requires a large number of emitters with a high spatial density as well as coherent optical transitions with small inhomogeneity. Here we show that by coupling nitrogen-vacancy (NV) centers in a diamond membrane to a high-finesse microcavity, also few, incoherent, inhomogeneous, and spatially separated emitters - as are typical for solid state systems - can enter the regime of collective emission. We observe a super-linear power dependence of the emission rate as a hallmark of collective emission. Furthermore, we find simultaneous photon bunching and antibunching on different timescales in the second-order auto-correlation function, revealing cavity-induced interference in the quantized emission from about fifteen emitters.
We develop theoretical models for mesoscopic emitter numbers to analyze the behavior in the Dicke state basis and find that the population of collective states together with cavity enhancement and filtering can explain the observations. Such a system has prospects for the generation of multi-photon quantum states, and for the preparation of entanglement in few-emitter systems.
\end{abstract}

\maketitle

\section{Introduction}

\begin{figure*}[ht]
    \centering
    \includegraphics[width=\textwidth]{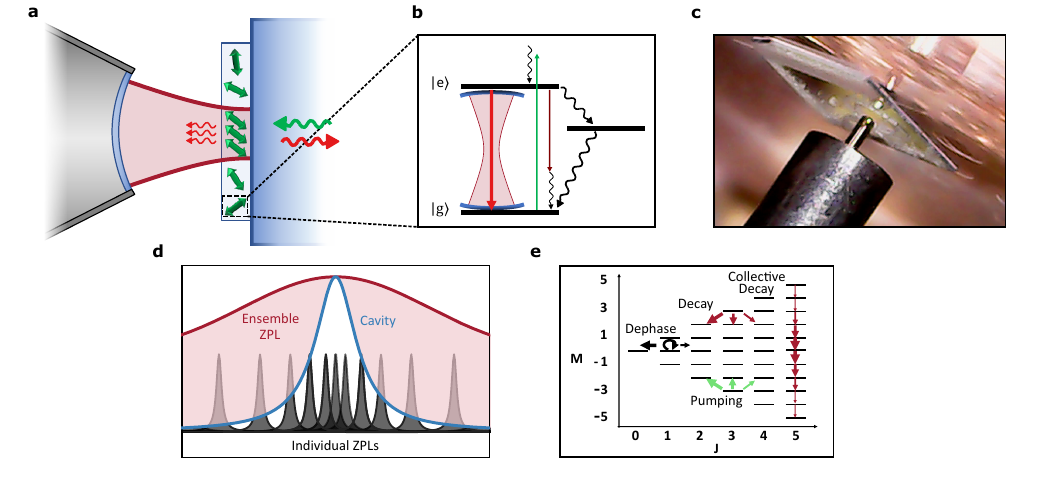}
    \caption[]{\textbf{Experimental setup and involved physics.} \textbf{a}, An NV center ensemble in a diamond membrane coupled to a fiber-based microcavity, where the NV centers are excited incoherently and emit collectively. \textbf{b}, Simplified level scheme of the NV center. The zero phonon line (ZPL) $\ket{g}\leftrightarrow \ket{e}$ is resonant with the cavity, leading to selective Purcell enhancement of the transition. The system is excited off-resonance via the phonon sideband (green arrow). \textbf{c}, Photograph of the cavity setup, showing the fiber micro mirror mounted in a steel needle, and the diamond membrane on a planar mirror. \textbf{d}, Schematic drawing of the inhomogeneously broadened ensemble ZPL, consisting of many individual transitions from which a few are overlapping with the narrow cavity resonance. \textbf{e}, Exemplary level scheme of collective Dicke states for ten identical two-level emitters where the quantum jumps due to the incoherent pumping (green arrows), dephasing (black arrows), individual and collective decay (red arrows) are indicated.}
    \label{fig:figure1}
\end{figure*}

Collective emission of quantum emitters is an intriguing behavior that can arise from the coupling of closely spaced atomic dipoles, where one of the most prominent features is a time-delayed superradiant pulse that grows quadratically in intensity with the emitter number \cite{dicke_coherence_1954}. It has been subject of numerous studies \cite{andreev_collective_1980,gross_superradiance_1982,raimond_statistics_1982,scully_super_2009,cong_dicke_2016,raino_superfluorescence_2018} and currently spurs great interest for possible applications, such as superradiant lasers \cite{meiser_prospects_2009,bohnet_steady-state_2012}, quantum memories \cite{ferioli_storage_2021,shen_subradiant_2022,lei_many-body_2022,rastogi_superradiance-mediated_2022} and quantum metrology \cite{paulisch_quantum_2019}.

Notably, even incoherent excitation can lead to spontaneous synchronization of emitters, an effect called superfluorescence (SF) \cite{bonifacio_quantum_1971,bonifacio_cooperative_1975}. This requires emitters with sufficient optical coherence and a high spatial density such that the average distance between two emitters is smaller than the transition wavelength. 
Reaching this regime can be challenging in a solid state environment, as emitters in dense ensembles typically suffer from effects like inhomogeneous and homogeneous broadening and spectral diffusion \cite{davies_jahn-teller_1981,tamarat_stark_2006,wolters_measurement_2013}. Recently, superfluorescence from perovskite quantum dot superlattices could be observed, which showed narrow ensemble emission lines under cryogenic conditions \cite{raino_superfluorescence_2018}. Surprisingly, also under strong phonon broadening at room temperature, high-density ensembles of nitrogen vacancy centers in nanodiamonds showed signatures of superfluorescence \cite{bradac_room-temperature_2017}. In both cases, a high density and a large number of emitters were crucial for the emergence of collective emission.
These requirements can be relaxed by coupling the emission to a high-finesse microcavity, which acts as a single-mode common light field that can couple also distant emitters and provides full spatial coherence \cite{Auffeves_few_2011,Leyman_2015}. Furthermore, the spontaneous emission rate is increased due to the Purcell effect \cite{purcell_resonance_1946}, and photons are funneled into a spectrally narrow cavity resonance \cite{grange_cavity-funneled_2015}, which improves their spectral coherence.

In an idealized picture, a system of $N$ identical emitters can be described in a basis of collective Dicke states $|J,M\rangle$ analogous to a number of spin $1/2$ systems that form a collective spin $J=0,1/2... N/2$ with a projection $M=-J ... J$ that represents the overall number of excitations $J+M$. As shown in Fig.~\ref{fig:figure1}\textbf{e}, the collective decay preserves $J$ and leads to quantum jumps between states with different $M$, while dephasing and individual decay cause quantum jumps also to states with different $J$. The relative rates of the jumps are illustrated by the thickness of the arrows, and their precise values can be found in Ref. \cite{Zhang_2018, Shammah_2018}. 
Overall, dephasing tends to reduce $J$ and lead to the population of subradiant states close to the lower boundary of the triangle. Due to their suppressed collective coupling to the cavity mode, these states promote excitation to higher levels under continuous pumping. These, in turn, can decay via the emission of multiple photons, leading to photon bunching \cite{Auffeves_few_2011,Bhatti_super_2015}, see Fig.~\ref{fig:figure1}\textbf{e}.
This has been observed for ultracold atoms \cite{hennrich_transition_2005}, and for larger numbers of solid state emitters \cite{Wiersig_Direct_2009,Jahnke_2016,bradac_room-temperature_2017,raino_superfluorescence_2018}.\\

Here, we report the observation of collective emission from a small ensemble of nitrogen-vacancy (NV) centers in diamond which display inhomogeneous and homogeneous broadening. They are dispersed over a few-micron large area in a diamond membrane and are resonantly coupled to a microcavity, see Fig.~\ref{fig:figure1}. Remarkably for such sample properties, we observe a super-linear increase of the emission rate with increasing incoherent excitation, indicating the onset of collective emission. On the other hand, we observe the hallmark of quantized emission of individual emitters with anti-bunching in the second-order correlation function. Furthermore, on a short time scale, we observe a sharp bunching peak. Notably, all these features are solely visible for the coherent zero-phonon-line (ZPL) photons at low temperature, while they are absent for the incoherent phonon side band (PSB) and at room temperature. To analyze the findings, we develop two complementary theoretical models that can treat the radiation of mesoscopic emitter numbers coupled to a cavity. Thereby we can reproduce the observations and obtain a detailed insight into the contribution of the population of collective Dicke states and the influence of the emitter number, dephasing, and inhomogeneity.

\section{Cavity-enhanced Fluorescence}

We incorporate NV centers in a thin membrane into a fully tunable, fiber-based microcavity with a quality factor of $Q=4.7\times 10^5$ at the wavelength of the ZPL of the NV centers, see Fig. \ref{fig:figure1}\textbf{a,b} and \textbf{c}. The cavity linewidth is much narrower than the inhomogeneously broadened ZPL of the ensemble (Fig. \ref{fig:figure1}\textbf{d}), such that a small number of resonantly coupling NV centers is spectrally selected. These emitters couple with a Purcell-enhanced rate to a single cavity mode.

\begin{figure*}[ht]
    \centering
    \includegraphics[width=\textwidth]{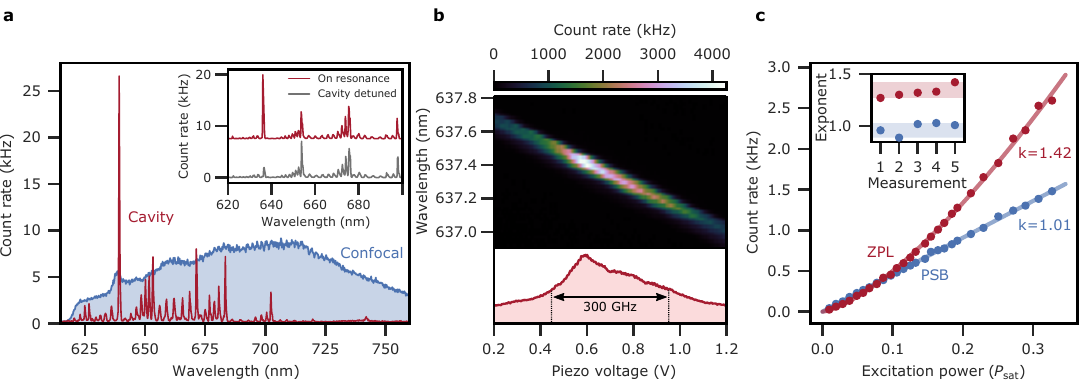}
    \caption[]{\textbf{Cavity-enhanced fluorescence of NV centers.} \textbf{a}, PL spectrum of an ensemble in the cavity (red) and in a confocal microscope (blue). The inset illustrates the decreased ZPL count rate as the cavity resonance is detuned from the transition by slightly changing the mirror separation. \textbf{b}, Cavity-enhanced spectroscopy of the ensemble ZPL, obtained by recording fluorescence spectra while step-wise changing the cavity length. The integrated counts reveal the inhomogeneous ZPL with a FWHM linewidth of $\delta \nu_\mathrm{inh}\approx 300$\,GHz. \textbf{c}, Comparison of ZPL and PSB fluorescence count rates, coupled to the same spatial cavity mode, in dependence of the excitation laser power. The ZPL features a super-linear increase, a characteristic of superfluorescent emission. The inset shows exponents obtained under different conditions like cavity length or lateral position.}
    \label{fig:figure2}
\end{figure*}

We study the cavity-enhanced fluorescence by recording photoluminescence (PL) spectra under off-resonant excitation. Therefore, the cavity fiber was laterally positioned such that the fundamental Gaussian mode shows an air-like character \cite{janitz_fabry-perot_2015} at $\lambda=\SI{637}{nm}$ with a finesse of $\mathcal{F}= 12000$. 
When tuning a cavity resonance to the ZPL, the resulting spectrum shows a series of sharp cavity resonances, with a dominant line at the ZPL wavelength (Fig.~\ref{fig:figure2}\textbf{a}). Within the spectrally broad phonon sideband, a large number of additional fundamental and higher order transverse cavity modes is visible. The measurement also illustrates the improved ratio of ZPL and phonon side band (PSB) photons, as compared to a confocal measurement at room temperature. This results from the Purcell effect, which leads to an increased emission rate and thereby an improved branching ratio, a higher emission directionality, leading to a better collection efficiency, and the suppression of off-resonant fluorescence by cavity spectral filtering. 
We remark that while for the cavity-enhanced ZPL emission, only about fifteen emitters contribute due to spectral filtering, all emitters (several hundred) spatially located within the cavity mode contribute to the resonances in the phonon side band, such that the ZPL enhancement is in fact more than an order of magnitude larger than is visible in Fig.~\ref{fig:figure2}\textbf{a}.
By applying a step-wise increasing voltage to the z-piezo and thus controlling the cavity length, the cavity resonance is tuned across the ensemble ZPL, see Fig. \ref{fig:figure2}\textbf{b}. Since the spectral cavity linewidth of $\delta \nu_\mathrm{cav}=\SI{1.0}{GHz}$ is much narrower than the inhomogeneously broadened ZPL, we can obtain a high-resolution spectrum. The summed counts reveal the spectral distribution of ZPL transitions, yielding a full width at half maximum (FWHM) of $\delta \nu_\mathrm{inh}\approx \SI{300}{GHz}$. When the cavity resonance frequency matches the center frequency of the ZPL, an up to 4-fold increase of fluorescence counts can be observed.

In a next step, we measure the power dependence of the fluorescence spectra under off-resonant excitation and compare the partially coherent emission into the ZPL with the red-shifted incoherent emission into the phonon side band. While the power dependence for the PSB emission matches the expected linear scaling when far from saturation, the trend for the ZPL emission clearly exhibits a super-linear behaviour following a power-law dependence $I\sim P^k$ with an exponent as high as $k=1.42$, see Fig.~\ref{fig:figure2}\textbf{c}. This is reproducible under various conditions like different cavity length or lateral position on the sample (inset of Fig.~\ref{fig:figure2}\textbf{c}) . The super-linear behaviour was only observed at cryogenic temperatures ($T=11$~K), in accordance with the prediction from a theoretical model (see below and SI). For an ensemble of superradiant emitters under pulsed excitation, one expects up to $k=2$. The lower value observed in the experiment can be explained by the inhomogeneity and the pure dephasing of the emitters, which leads to a population of Dicke states towards low $J$ values. These states decay at a lower rate, which corresponds to a smaller value for $k$.

\begin{figure*}[ht]
    \centering
    \includegraphics[width=0.9\textwidth]{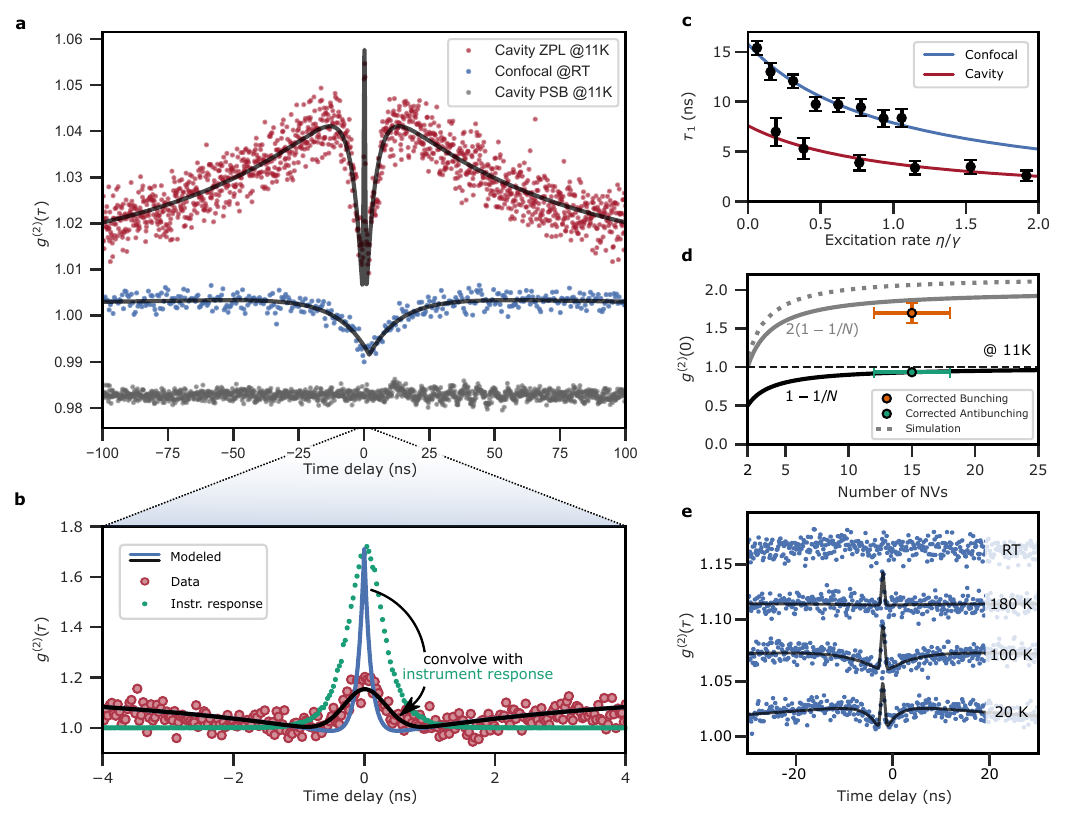}
    \caption{\textbf{Temporal photon statistics of NV fluorescence.} \textbf{a}, Second-order autocorrelation function of the emitted photons. The cavity enhanced fluorescence (red data points) reveals bunching on a sub-ns timescale, as opposed to the fluorescence recorded in a confocal microscope (blue data points). The PSB emission in the cavity setup (grey data points, shifted by 0.02 for clarity) does not show any bunching or antibunching. \textbf{b}, Enlarged view of the bunching around $\tau=0$ together with the measured instrument response function. A modeled $g^{(2)}(\tau)$ function is convolved with the instrument response to show the agreement with the background-corrected experimental data. \textbf{c}, Antibunching time constants for different excitation laser intensities, extracted from the respective correlation functions. The fits correspond to a rate equation model, and allow one to infer the effective Purcell factor $F_{P,eff}=1.1\pm0.3$. \textbf{d}, From the measured antibunching amplitude, one can extract the number of contributing emitters $N=15\pm 3$. For $N$ individually decaying emitters, one expects a thermal bunching amplitude of $2(1-1/N)$ (solid grey line), which coarsely agrees with our measured value (orange data point). The prediction of our model is shown for comparison (dashed grey line). \textbf{e}, A set of $g^{(2)}$ functions for increasing temperature reveals that the bunching amplitude is robust against further dephasing.}
    \label{fig:figure3}
\end{figure*}

\section{Photon statistics}

Further insight can be obtained by studying the photon statistics of the radiation emitted from the cavity. Therefore, we use a Hanbury-Brown and Twiss (HBT) setup and record the second-order autocorrelation function $g^{(2)}(\tau)$.
An exemplary $g^{(2)}(\tau)$ of the cavity-enhanced ZPL fluorescence is shown in Fig. \ref{fig:figure3}\textbf{a}, where three contributions can be fitted by the function
\begin{align}
    g^{(2)}(\tau)=1-\left(a+b\right)\,e^{-\frac{|\tau|}{\tau_1}}+b\,e^{-\frac{|\tau|}{\tau_2}}+\phi_{c,\tau_3}(\tau).
\end{align}
Here, $a$ and $\tau_1$ ($b$ and $\tau_2$) describe the single-emitter antibunching (bunching) amplitude and time constant that are linked to the NV center excited (shelving) state lifetime.

We can extract the cavity-shortened excited state lifetime and thus the Purcell factor from the antibunching time constant $\tau_1(I)$, which depends on the excitation rate $\eta$ and thus the laser intensity $I$. Therefore, we perform a power-dependent set of measurements of the $g^{(2)}$-function and evaluate $\tau_1(I)$, see Fig.~\ref{fig:figure3}\textbf{c}. Extrapolating to zero excitation intensity yields the excited state lifetime. The fit function $\tau_1(I)=\tau_1(0)/\left(1+\sigma I\tau_1(0)\right)$ can be deduced from simplified two-level rate equations, where $\sigma I=\eta$ is the excitation rate from the ground to the excited state. With this analysis, the excited state lifetime of the cavity-coupled NV center ensemble was estimated as $\tau_\mathrm{c}=7.6\pm 0.9$\,ns. A second set of power-dependent $g^{(2)}$ functions was taken in a confocal microscope, from which the excited state lifetime without cavity-enhancement was estimated as $\tau_\mathrm{0}=15.8\pm 0.8$\,ns. The increased lifetime compared to the bulk value of 13 ns can be explained by the presence of the nearby diamond-silica interface for shallowly implanted NV centers as studied here, which leads to a modified local dielectric environment that reduces the decay rate \cite{radtke_nanoscale_2019,zahedian_modeling_2023}. From the ratio of the cavity-enhanced and free-space lifetime, we can extract the effective Purcell factor to be $F_{\mathrm{P,eff}}=\tau_\mathrm{0}/\tau_\mathrm{c}-1=1.1\pm 0.3$, which corresponds to an ideal Purcell factor $F_{\mathrm{P,ideal}}=F_{\mathrm{P,eff}}/\xi=36\pm 11$ with the Debye-Waller factor $\xi=0.03$ for NV centers \cite{faraon_resonant_2011}. This value is in the range of expected Purcell factors for the given cavity finesse, mode volume, dipole orientation and random spectral cavity-emitter overlap.\\
From the amplitude of the antibunching dip, we can deduce the number of emitters that couple to the cavity and thus dominantly contribute to the ZPL signal. Therefore, we follow the prediction for $N$ independent emitters, which is given by $g^{(2)}(0)=1-1/N$. The antibunching contrast $a$ is furthermore reduced by the purity $p$, which we define as the fraction of the collected cavity radiation that originates from ZPL emission. Since a large number of spectrally detuned NV centers additionally contribute to the collected cavity resonance via their phonon sideband, they introduce uncorrelated background and thus reduce the purity. We can infer $p$ by measuring the cavity-collected PSB background when detuning the cavity from the ensemble ZPL (see Fig.~\ref{fig:figure2}\textbf{b}) and obtain $p = 0.5$. We use $a/p^2=1/N$ to obtain $N$.
In the case of the measurement shown in Fig.~\ref{fig:figure3}\textbf{a}, this analysis yields $N=15\pm3$ emitters coupling to the cavity.
For comparison, we observe a flat $\mathrm{g^{(2)}}$ function when selectively collecting the cavity-coupled PSB emission, and an antibunching contrast of $a\approx0.01$ in a measurement of the same sample region with a confocal microscope at room temperature. The latter contrast is smaller compared to the cavity ZPL measurement since no spectral filtering is present and all excited emitters contribute. However, the spatial excitation and collection modes of the confocal microscope are smaller than the cavity mode cross section, such that fewer emitters are contributing than in the cavity under PSB collection. The antibunching contrast yields an estimate of hundred emitters in the confocal point-spread function and - scaled to the mode area of the cavity - several hundred emitters in the spatial cavity mode. This relatively high number explains why we do not observe any $\mathrm{g^{(2)}}$-contrast in the cavity-filtered PSB.

We finally turn to the additional bunching peak close to zero time delay. We observe a Gaussian shape described with $\phi_{c,\tau_3}(\tau)$ in Eq.~1 with a FWHM of $\approx 400$\,ps. This time constant is entirely limited by the instrument response of the HBT setup, originating from the timing jitter of the APDs. A measurement with a pulsed laser is shown in Fig.~\ref{fig:figure3}\textbf{b}. In the limit of weak absorption, the fastest possible time scale for emission from the cavity is given by the cavity intensity decay time of $T_{cav}=\delta \nu_{cav}^{-1}=100$\,ps. This value is also in accordance with our model considering emitters with additional dephasing that couple collectively to a cavity mode (see Fig.~\ref{fig:figure4}\textbf{c} and SI). Correcting for the incoherent background from uncoupled NV centers and deconvolving the measured bunching peak allows us to estimate the intrinsic peak value of the bunching (Fig.~\ref{fig:figure3}\textbf{b}). We obtain a value of $g^{(2)}(0)=1.7$, which is close to an analytical estimate for interfering independent emitters \cite{Auffeves_few_2011}, $g^{(2)}(0)=2(1-1/N)$ for our extracted emitter number $N=15\pm 3$. This thermal limit describes individually decaying but identical emitters and neglects possible contributions from collective states. We compare the prediction for $g^{(2)}(0)$ from our model to the analytical expression in Fig.~\ref{fig:figure3}\textbf{d}, and only a minor difference is visible.
In contrast, no bunching is observed in the cavity-enhanced PSB at any temperature, as well as in the confocal measurement at room temperature.

Finally, to study the effect of dephasing more explicitly, we perform measurements of the $g^{(2)}$ function as a function of temperature (Fig.~\ref{fig:figure3}\textbf{e}). We observe a decrease of the antibunching amplitude, in accordance with the prediction of our model when assuming a temperature-dependent dephasing rate as measured in \cite{fu_observation_2009}. We note that at elevated temperature, also an increased number of emitters are expected to contribute to the signal. This is due to the increased homogeneous linewidth, which leads to a growing spectral overlap of detuned emitters with the cavity mode. The temperature-dependent collective bunching amplitude remains approximately constant, even when approaching room-temperature. However, when accounting for the changing background contribution, an overall decrease results.

The measurements show that while the bunching contrast is not very sensitive to collective effects in the presence of dephasing, the observed signatures can not be described solely by individually decaying emitters. Together with the super-linear power dependence of the cavity emission, the observations support the picture that the Purcell-enhanced emission into a single cavity mode synchronizes emission if strong enough cavity coupling and sufficient coherence is present.

\section{Numerical Model}
\begin{figure*}[ht]
    \centering
    \includegraphics[width=0.9\textwidth]{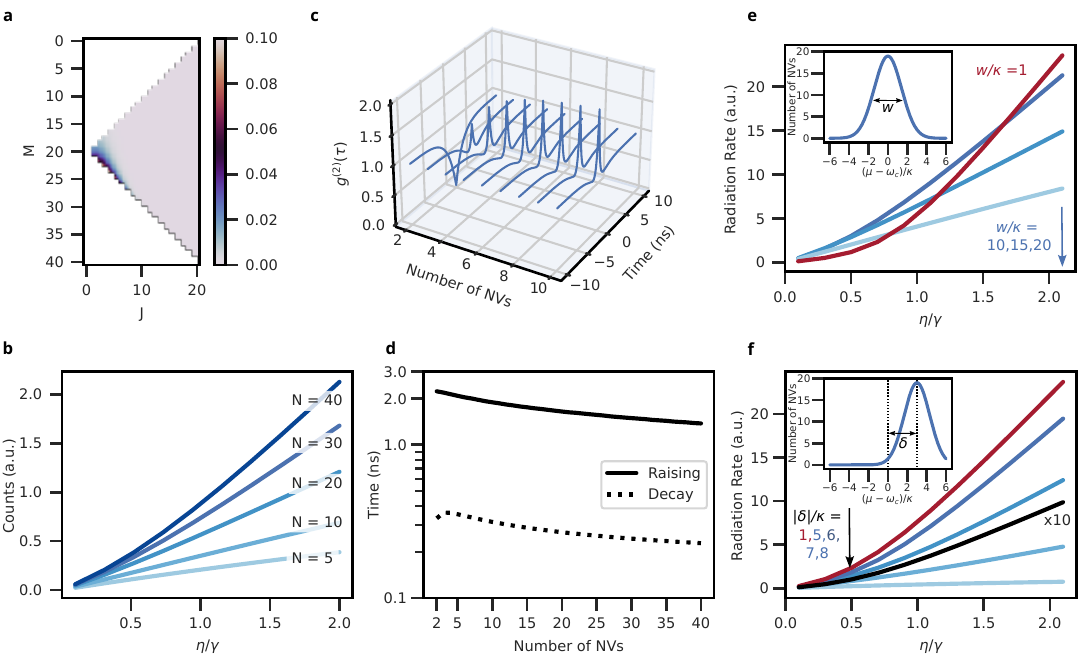}
    \caption{\textbf{Modeling of collective emission.} \textbf{a-d} Results for an ensemble with identical NV centers. \textbf{a} Population of Dicke states under continuous incoherent pumping for 40 NV centers. While population peaks at the lower-left boundary of the triangle, also higher-lying collective states are populated. \textbf{b}, Radiation rate as a function of the ratio $\eta/\gamma$ between the pumping and decay rate, shown for increasing number of NV centers $N$. \textbf{c} Second-order correlation function $g^{(2)}(\tau)$ for different values of $N$ and $\eta/\gamma=2$. \textbf{d} Corresponding decay and rise time of $g^{(2)}(\tau)$. \textbf{e,f}, Results for NV centers with different frequencies (sub-ensembles). \textbf{e}, Influence of a Gaussian distributed frequency inhomogeneity of 500 NV centers with a width $w$ on the radiation rate. \textbf{f}, Influence of the frequency detuning $\delta$ on the radiation rate, and the radiation weighted by the frequency fluctuation (black line, amplified by 10 times), treated by a Gaussian distribution of $\delta$ with a width $300$~GHz.}
    \label{fig:figure4}
\end{figure*}
For a deeper analysis, we develop numerical models for small ensembles of emitters coupled to a cavity.
We study first the case of identical NV centers to explain the main physics (Fig.~\ref{fig:figure4}\textbf{a-d}), and then we introduce sub-ensembles of NV centers with different transition frequencies to examine the influence of the frequency inhomogeneity and fluctuations (Fig.~\ref{fig:figure4}\textbf{e,f}). For the sake of simplicity, we focus here on the transition between the $m_s=0$ levels of the triplet states $^3A_2, ^3E$, because it dominates the radiation under continuous illumination. In our theoretical model (see Methods section), we treat this transition as two-level system in the presence of decay, dephasing, incoherent pumping with rates $\gamma, \chi,\eta$ and coupled by a strength $g$ to an optical cavity with a frequency $\omega_c$ and a loss rate $\kappa$. All these processes are described by a quantum master equation (QME). 
For the homogeneous system, all the NV centers are identical and they can be well described with Dicke states $|J,M\rangle$, and the QME can be solved in the basis of these states \cite{Zhang_2018, Shammah_2018}.
Under the incoherent pumping with rate $\eta$, the excited ensemble emits radiation. Since our system is in the weak coupling regime, we can eliminate adiabatically the cavity to obtain the radiation rate $I_\mathrm{rad}=\Gamma \sum_{J,M}|A_{J,M}^-|^2\rho_{MM}^J$, where $\Gamma\approx 4g^2/\kappa$ is the cavity-mediated Purcell-enhanced decay, the coefficient $A_{JM}^-=\sqrt{(J+M)(J-M+1)}$ arises from the action of the collective lowering operator, and $\rho_{MM}^J$ is the total population of the degenerate Dicke state manifold with same quantum numbers $J$ and $M$. As $\eta$ increases, the steady state of the ensemble rises along the lower boundary of the Dicke states with decreasing values of $J$ and $M=-J, -J+1,-J+2,...$ (Fig.~\ref{fig:figure4}\textbf{a}), which leads to the increased radiation. Furthermore, as the number of NV centers increases, the number of excitations in the system increases, and the scaling of radiation changes from linear to super-linear (Fig.~\ref{fig:figure4}\textbf{b}). 
By applying the quantum regression theorem, we have also computed the second-order auto-correlation function $g^{(2)}(\tau)$. In general, this function does not change qualitatively with increasing pumping rate (Fig. S3 of SI), and thus, here, we focus on the influence of the number of NV centers $N$ for the maximal pumping rate $\eta=2\gamma$ in Fig.~\ref{fig:figure4}\textbf{c}. While the $g^{(2)}(\tau)$ function shows a pronounced dip for a single NV center, it shows a sharp bunching peak around zero delay time $\tau=0$ for $N\geq 2$, and an anti-bunching dip for longer time. The quantitative analysis of the time constants is illustrated in Fig.~\ref{fig:figure4}\textbf{d}. The decay time of the peaks (the rising time of the dips) reduces to about 0.23 ns (1.37 ns) with increasing $N$. Note that the decay time increases slightly when $N$ changes from 2 to 3. For $\eta=2\gamma$ and $N=40$, the Dicke state $|J=4,M=-4\rangle$ has the largest population (Fig.~\ref{fig:figure4}\textbf{a}), and the singly (doubly) excited state $|J=4,M=-3(-2)\rangle$ has a decay rate of $\Gamma |A_{(4,-3)}^-|^2=8\Gamma, (\Gamma |A_{(4,-2)}^-|^2=18\Gamma)$. Since the inverse of these rates $1/(8\Gamma)\approx0.33$~ns and $1/(18\Gamma)\approx0.15$~ns are close to the decay time 0.23 ns, the sharp anti-bunching peak of the $g^{(2)}(\tau)$ function should be associated with the populated Dicke states. When introducing additional dephasing at a value equal to the cavity decay rate, we observe a reduced bunching peak time constant of $110$~ps, closely matching the cavity decay rate of $100$~ps, (see SI). This motivates the choice of the time constant in the model for Fig.~\ref{fig:figure3}b. At the same time, the peak maximum (dip minimum) increases and saturates gradually to a value slightly above 2 (below 1) for larger $N$. While the dip minimum matches closely with the analytical prediction, the bunching peak remains systematically higher than given by the analytical expression and is shown in Fig.\ref{fig:figure3}\textbf{d}. With the above results, we have confirmed that the non-linear scaling and the sharp bunching peaks, as observed in the experiment, are caused by the cavity-mediated collective radiation of the NV centers.

To reflect more closely the experimental conditions, we study the influence of frequency inhomogeneity and fluctuations \cite{Bychek_superradiant_2023}. We divide the total NV centers into multiple sub-ensembles with different frequencies, and assume that the number of NVs as function of the transition frequency follows a Gaussian distribution with a mean $\mu$ and a linewidth $w=2\sqrt{2\mathrm{ln}(2)} \sigma$ (standard deviation $\sigma$). Although we cannot solve the QME of such system with the density matrix technique due to the exponentially increased Hilbert space, we can solve it with the second-order mean-field theory, and derive semi-analytical expressions for the quantities characterizing the system steady-state. In our simulations, we consider 500 NV centers, as estimated from the experimentally measured $g^{(2)}$ function. If the Gaussian distribution is resonant with the cavity and has a similar width $(\mu=\omega_c,w=\kappa)$, the radiation rate shows obvious super-linear and linear scaling for $\eta/\gamma \leq 1.2$ and $\eta/\gamma > 1.2$, respectively (Fig.~\ref{fig:figure4}\textbf{e}). However, as $w$ increases to $5\kappa$, the super-linear scaling becomes less pronounced. If $w$ increases further to $20\kappa$, the radiation rate decreases, and approaches gradually the linear scaling. If the Gaussian distribution with a fixed width $w=3\kappa$ is detuned from the cavity by an increasing amount $\delta = \mu-\omega_c$, the radiation rate reduces slightly for smaller detuning $|\delta|\leq 3 \kappa$, but dramatically for much larger detuning $|\delta| > 3 \kappa$ (Fig.~\ref{fig:figure4}\textbf{f}). Furthermore, by assuming that the probability of the detuning $\delta$ follows also a Gaussian distribution with a large linewidth $w\approx 300$~GHz, as estimated from Fig.~\ref{fig:figure2}\textbf{b}, we can weigh the above results with this probability to obtain the averaged radiation (black line in Fig.~\ref{fig:figure4}\textbf{f}), and observe the persistence of the super-linear scaling. 
To evaluate the influence of the frequency inhomogeneity and fluctuation on the $g^{(2)}$ function, we adopted an indirect approach, and verify firstly that the radiation is affected by dephasing of NV centers in a similar way as by the frequency inhomogeneity (see SI), which permits us to account for the inhomogeneity in the model of identical NV centers with a single dephasing rate \cite{Wu_2021,Julsgaard_2012}. The sharp bunching peak of the $g^2$-function is preserved in the presence of large dephasing and frequency fluctuation, which is accord with the fact that such behavior appears already for the system with as few as two NV centers (Fig.~\ref{fig:figure1}\textbf{c}). In the end, we note that the observed bunching shoulder of the $g^{(2)}(\tau)$ function at longer times is known to originate from the singlet excited levels $^1A_1, ^1E$ of NV centers, and thus improved models including also these levels should be able to capture all these characteristics. Due to the complexity of the models, we defer their discussion to a more detailed theoretical paper.

\section{Conclusions}
In summary, we demonstrate that incoherent, inhomogeneous and spatially separated emitters can yield signatures of collective emission, enabled by the coupling to a common cavity mode. We have focused on the nonlinear scaling and the photon bunching of the incoherently pumped NV centers. It would be interesting to further explore the collective radiation from population-inverted NV centers, or from a system with more NV centers, which might enter the strong coupling regime.
Cavity-induced synchronization can furthermore lead to non-trivial collective behavior even for emitters outside the cavity linewidth \cite{Debnath_2019}. Collective emission may be used to generate multi-photon quantum states \cite{munoz_emitters_2014} and can be a promising route to produce multi-qubit entangled states \cite{Santos_2022,Roulet_2018}. It will be an intriguing direction to explore cooperative behavior in the cavity quantum electrodynamics regime that has recently been demonstrated for two emitters in cavities \cite{Reimann_2015,Neuzner_2016,Kim_2018,evans_photon-mediated_2018,lukin_two_2023} and expand it to a controlled number of few emitters. 

\section*{Appendix A: Experimental setup}
The sample used in this experiment is an electronic grade single crystal CVD-grown diamond with a [100] surface orientation. The sample was initially cut to have an average thickness of $40\pm10$\,\textmu m and a thickness variation of $<1$\,\textmu m/mm. Afterwards, the sample was shallowly implanted by the manufacturer with nitrogen ions, leading to a high density of NV centers ($\sim$300\,NV/\textmu m$^2$) at a depth of 10\,nm. The sample was further cut to have a footprint of $1.5\times1.5$\,mm$^2$ and bonded onto a dielectric mirror by van der Waals bonding. Afterwards, the diamond was machined by inductively-coupled plasma reactive ion etching, creating a membrane with a minimal thickness of 3.5\,\textmu m and a surface roughness as low as 0.3\,nm\,rms in an area of $4\times4$\,\textmu m$^2$ \cite{Heupel_2020}.

The sample is integrated into a fiber-based microcavity, consisting of a macroscopic plane mirror which holds the diamond and a laser-machined fiber mirror.
We excite the NV centers incoherently using a continuous wave laser with $\lambda_{exc}=532$\,nm through the free space port of the cryostat. The same optical path is used to filter and collect fluorescence light.

A more detailed description of the optical setup can be found in the SI.

\section*{Appendix B: Theoretical model}
For the system with multiple NV sub-ensembles, the dynamics is described by the quantum master equation (QME): 
\begin{equation}
\begin{split}
    &\frac{\partial}{\partial t}\hat{\rho}=-\frac{i}{\hbar}\left[\hat{H}_{tls}+\hat{H}_c+\hat{H}_{tls-c},\hat{\rho}\right]-\kappa \mathcal{D}[\hat{a}]\hat{\rho}\\\
    &- \sum_{k=1}^{N_{ene}}\sum_{i=1}^{N_k}\left\{ \eta_k\mathcal{D}[\hat{\sigma}_{k,i}^+]\hat{\rho} + \gamma_k\mathcal{D}[\hat{\sigma}_{k,i}^-]\hat{\rho} + \frac{\chi_k}{2}\mathcal{D}[\hat{\sigma}_{k,i}^z]\hat{\rho} \right\}
\end{split}
\end{equation}
The Hamiltonian $\hat{H}_{tls}=\hbar\sum_{k=1}^{N_{ene}}(\omega_k/2)\sum_{i=1}^{N_k}\hat{\sigma}_{k,i}^z$ describes the NV centers as two-level systems with the transition frequency $\omega_k$ and the Pauli operator $\hat{\sigma}_{k,i}^z$, where the labels $k,i$ indicate the individual sub-ensemble and the individual NV center. The Hamiltonian $\hat{H}_c=\hbar\omega_c\hat{a}^+\hat{a}$ describes the optical cavity with frequency $\omega_c$, the photon creation $\hat{a}^+$ and annihilation operators $\hat{a}$. The Hamiltonian $\hat{H}_{NV-c}=\hbar\sum_{k=1}^{N_{ene}}g_k\sum_{i=1}^{N_k}\left\{\hat{\sigma}_{k,i}^+\hat{a}+\hat{a}^+\hat{\sigma}_{k,i}^-\right\}$ describes the NVs-optical cavity coupling with a strength $g_k$, and the raising and lowering operators $\hat{\sigma}_{k,i}^+$, $\hat{\sigma}_{k,i}^-$. The Lindblad terms describe the cavity photon loss, the individual decay, dephasing, pumping of the two-level systems with rates $\kappa$, $\gamma_k$, $\chi_k$, $\eta_k$, respectively, and the superoperator is defined as  $\mathcal{D}[\hat{o}]\hat{\rho}=\frac{1}{2}\left\{\hat{o}^+\hat{o},\hat{\rho}\right\}-\hat{o}\hat{\rho}\hat{o}^+$ (for any operator $\hat{o}$). The second order auto-correlation function can be computed with the expression $g^{(2)}(\tau)=\braket{\hat{a}^+(t_{ss})\hat{a}^+(t_{ss}+\tau)\hat{a}^-(t_{ss}+\tau)\hat{a}^-(t_{ss})}/\braket{\hat{a}^+(t_{ss})\hat{a}^-(t_{ss})}$, where $t_{ss}$ refers to the time for the system reaching steady-state. The numerator can be transformed into the expression $\mathrm{tr}\left\{\hat{a}^+\hat{a}^-\hat{\varrho}(\tau)\right\}$ in the Schrödinger picture, where the ancillary operator $\hat{\varrho}(\tau)$ satisfies also the QME as $\hat{\rho}(\tau)$ with however the initial condition $\hat{\varrho}(0)=\hat{a}^-\hat{\rho}(t_{ss})\hat{a}^+$ according to quantum regression theorem.

To describe the ideal system with identical NV centers, we remove the label \(k\) in the above model, and further eliminate adiabatically the optical cavity to achieve an effective master equation, which differs from the original one by replacing \(\hat{H}_c + \hat{H}_{tls-c}\) with \(\hat{H}_{shift} = \hbar(\delta\omega/2) \sum_{i=1}^{N} \hat{\sigma}_i^z\), and \(\kappa \mathcal{D}[\hat{a}] \hat{\rho}\) with \(\Gamma \mathcal{D}[\sum_{i=1}^{N} \hat{\sigma}_i^-] \hat{\rho}\), where \(\delta\omega = 2g^2 (\omega_c - \omega_0)/((\omega_c - \omega_0)^2 + \left((\kappa+\eta+\gamma)/2+\chi\right)^2)\) and \(\Gamma=g^2\kappa/((\omega_c - \omega_0)^2 + \left((\kappa+\eta+\gamma)/2+\chi\right)^2)\) are the cavity-induced frequency shift and Purcell-enhanced decay rate. To proceed, we introduce the collective operators \(\hat{J}_x = \frac{1}{2} \sum_{i=1}^{N} (\hat{\sigma}_i^- + \hat{\sigma}_i^+)\), \(\hat{J}_y = \frac{i}{2} \sum_{i=1}^{N} (\hat{\sigma}_i^- - \hat{\sigma}_i^+)\), \(\hat{J}_z = \frac{1}{2} \sum_{i=1}^{N} \hat{\sigma}_i^z\), \(\hat{J}^2 = \sum_{i=x,y,z} (\hat{J}_i)^2\), and define the Dicke states from the eigen-value problems \(\hat{J}^2 |J,M\rangle = J(J+1)|J,M\rangle\) and \(\hat{J}_z |J,M\rangle = M|J,M\rangle\). By expressing the Dicke states for \(N\) two-level systems as the product states of \(N-1\) two-level systems and one two-level system (using the Clebsch-Gordan coefficients), one can identify the dissipative effects in the collective Dicke state picture induced by the incoherent pumping, decay, and dephasing of individual two-level systems \cite{Zhang_2018,Shammah_2018} (Fig.~\ref{fig:figure1}\textbf{e}). In this way, we can derive and then solve the equation for the density matrix elements \(\rho_{MM'}^J(t)\) within the Dicke states space (the density matrix is diagonal in \(J\)). From \(\rho_{MM'}^J(t)\), we can extract the population of the Dicke states \(P_{JM}(t) = \rho_{MM}^J(t)\), the radiation rate \(I_{rad}(t) = \Gamma\langle\hat{J}^+\hat{J}^-\rangle(t) = \Gamma \sum_{J,M}|A_{JM}^-|^2 P_{JM}(t)\), and the second-order auto-correlation function \(g^{(2)}(\tau) = \langle\hat{J}^+(t_{ss})\hat{J}^+(t_{ss}+\tau)\hat{J}^-(t_{ss}+\tau)\hat{J}^-(t_{ss})\rangle / (I_{rad}(t_{ss})/\Gamma)^2\), where the numerator can be calculated in a similar way as stated in the previous paragraph. In Sec. S1 of the Supporting Information, we provide the corresponding codes based on the QuTip package \cite{Julsgaard_2012}.

For the system with multiple NV sub-ensembles, we adopt a second-order mean-field approach to derive the equation
\(
\frac{\partial}{\partial t} \langle\hat{o}\rangle = \text{tr}\{\hat{o}(\frac{\partial}{\partial t} \hat{\rho})\}
\)
for the mean value
\(
\langle\hat{o}\rangle = \text{tr}\{\hat{o}\hat{\rho}\}
\)
from the QME, apply the approximation
\(
\langle\hat{o}\hat{p}\hat{q}\rangle \approx \langle\hat{o}\rangle\langle\hat{p}\hat{q}\rangle + \langle\hat{p}\rangle\langle\hat{o}\hat{q}\rangle + \langle\hat{q}\rangle\langle\hat{o}\hat{p}\rangle - 2\langle\hat{o}\rangle\langle\hat{p}\rangle\langle\hat{q}\rangle
\)
to truncate the equation hierarchy to obtain a closed set of equations and assume the NV centers in each sub-ensemble to be identical. Furthermore, we consider the equations at the steady-state, and derive the semi-analytical expressions for the intracavity photon number
\(
\langle\hat{a}^\dagger \hat{a}\rangle \approx i\sum_{k=1}^{N_{\text{ene}}}(N_k g_k/\kappa)(\langle\hat{\sigma}_{ki}^+ \hat{a}\rangle - \langle\hat{a}^\dagger \hat{\sigma}_{ki}^+\rangle)
\)
, the population differences
\(
\langle\sigma_{ki}^z\rangle \approx \left[-2ig_k (\langle\sigma_{ki}^+ \hat{a}\rangle-\langle\hat{a}^\dagger \sigma_{ki}^+\rangle) + \eta_k - \gamma_k\right]/(\gamma_k + \eta_k),
\)
and the NV-NV correlations
\(
\langle\sigma_{ki}^+ \sigma_{k'i'}^-\rangle \approx (\omega_k^* - \omega_{k'})^{-1}(g_k \langle\sigma_{ki}^z\rangle \langle\sigma_{k'i'}^+ \hat{a}\rangle^* - g_{k'} \langle\sigma_{k'i'}^z\rangle \langle\sigma_{ki}^+ \hat{a}\rangle)
\)
and the NV-photon correlations
\(
\langle\sigma_{ki}^+ \hat{a}\rangle \approx (\omega_{ki}^* - \omega_c)^{-1} [g_k (\langle\hat{a}^\dagger \hat{a}\rangle \langle\sigma_{ki}^z\rangle - \langle\sigma_{ki}^+ \sigma_{ki'}^-\rangle) + \sum_{k'\neq k}^{N_{\text{ene}}}N_{k'}g_{k'} \langle\sigma_{ki}^+ \sigma_{k'i'}^-\rangle + \frac{g_k}{2}(\langle\sigma_{ki}^z\rangle + 1).
\)
By taking the complex conjugate of the last equation, we can also obtain the expressions for
\(
\langle\hat{a}^\dagger \sigma_{ki}^+\rangle.
\)
In these expressions, the complex frequencies are defined as
\(
\omega_k=\omega_k-i(\gamma_k+\eta_k)+\chi_k, \quad \omega_c=\omega_c-i\kappa/2.
\)
In practice, we insert the first three expressions into the last one, and then solve the self-consistent equations for the NV-photon correlations with the FindRoot command in Mathematica; see the codes in Sec. S1 of the SI.


\end{document}